# The Black Hole Final State

Gary T. Horowitz[1] and Juan Maldacena,[2]

[1] *University of California at Santa Barbara*
*Santa Barbara, CA 93106, USA*

[2] *Institute for Advanced Study*
*Princeton, New Jersey 08540, USA*

We propose that in quantum gravity one needs to impose a final state boundary condition at black hole singularities. This resolves the apparent contradiction between string theory and semiclassical arguments over whether black hole evaporation is unitary.

## 1. Introduction

There is some evidence that in string theory the formation and evaporation of black holes is a unitary process. This evidence comes mainly from conjectured non-perturbative descriptions of string theory in asymptotically flat [1] or asymptotically anti de Sitter (AdS) [2,3,4] spacetimes. However semiclassical arguments suggest that this process is not unitary and causes pure states to evolve into mixed states [5]. A natural question is what goes wrong with the semiclassical reasoning.

The purpose of this note is to provide a possible answer to this question. Rather than the radical modification of quantum mechanics required for pure states to evolve into mixed states, we adopt a more mild modification. We propose that at the black hole singularity one needs to impose a unique final state boundary condition. More precisely, we have a unique final wavefunction for the interior of the black hole. Modifications of quantum mechanics where one imposes final state boundary conditions were considered in [6,7,8,9]. Here we are putting a final state boundary condition on part of the system, the interior of the black hole. This final boundary condition makes sure that no information is "absorbed" by the singularity.

This proposal naturally resolves the black hole information puzzle in the following way. In the process of black hole evaporation, particles are created in correlated pairs with one falling into the black hole and the other radiated to infinity. The correlations remain even when the particles are widely separated. The final state boundary condition at the black hole singularity acts like a measurement that collapses the state into one associated with the infalling matter. This transfers the information to the outgoing Hawking radiation in a process similar to "quantum teleportation" [10].[1]

Since the final state is fixed, one might wonder if infalling observers will feel the arrow of time flowing backwards. We argue that this does not occur due to the following two observations. First, we assume that the final boundary state is a highly entangled, non-local combination of the states of the infalling matter and the states of the infalling Hawking quanta. Second, an infalling observer has access to only a small fraction of the degrees of freedom of the system. So when the infalling observer performs any measurement she will have to trace over the unobserved states. This will effectively allow many possible final states.

---

[1] It is different than standard quantum teleportation because the uniqueness of the final state makes it possible to transfer the quantum information without telling the outside observer the result of measurements in the interior.

Our proposal is closely connected with the idea that the wavefunction of the universe is unique [11]. If there is a unique quantum state associated with the big bang singularity, it is not implausible that there is a unique quantum state associated with the black hole singularity. Unfortunately we are not able to give a precise method for computing this wavefunction. It should be determined by a full quantum theory of gravity. However, we point out that euclidean AdS/CFT suggests the existence of a unique wavefunction of the universe. In fact, the field theory gives a mathematically precise way of defining the Hartle-Hawking no-boundary state in a special asymptotic region.

In the next section we review semiclassical black hole evaporation and previous work on quantum mechanics with final state boundary conditions. Our proposal is discussed in section 3, including arguments that time will not appear to run backwards inside a black hole, and the relation to black hole complementarity. In Section 4 we describe some generalizations to other types of singularities, including charged black holes and cosmological singularities. We also discuss the implications of euclidean AdS/CFT for the wavefunction of the universe. The final section contains some concluding remarks.

## 2. Review of previous results

### 2.1. Review of semi-classical black hole evaporation

We begin by reviewing Hawking's computation of black hole evaporation [12]. For simplicity, we will consider gravity coupled to a single scalar field $\phi$. One starts with a classical solution describing the scalar field collapsing and forming a black hole. Then one quantizes the fluctuations in $g_{\mu\nu}$ and $\phi$ about this solution. These fluctuations start in the vacuum state in the far past. This vacuum state evolves in a time dependent fashion due to the time dependence of the metric. At late times we can choose a spacelike slice that goes through the horizon. The horizon divides this spacelike hypersurface into two parts, one inside and one outside the horizon. We can separate the Hilbert space of the fluctuations into two parts $H_{in}$ and $H_{out}$ which contain wavefunctions localized inside or outside the horizon. The original vacuum state evolves to a state, called the Unruh state $|U\rangle$ [13], which lives in $H_{in} \otimes H_{out}$. It turns out that this state contains a flux of outgoing particles in $H_{out}$. At this point unitarity is still preserved. However, for the outside observer, unitarity is effectively lost since he will not be able to do any measurement in $H_{in}$ [5]. So he will be forced to average over the unobserved states in the interior, thus obtaining a density matrix in $H_{out}$. More importantly, when the backreaction is included in a semi-classical approximation, the black hole will slowly lose all its mass through Hawking radiation and disappear. From the classical black hole geometry, the information about what formed the black hole cannot come out without violating causality. So at late times, unitarity is violated even in principle since the part of the state behind the horizon is gone.

For what we are going to do later it is convenient to introduce some notation. We will be interested in considering different infalling matter configurations. Even though they are treated classically in the usual semiclassical description, we will view them as states in a Hilbert space $H_M$. In addition we have the Hilbert space $H_{in}$ and $H_{out}$ introduced above. The semi-classical definition of these Hilbert spaces depends on the initial matter state via the black hole mass $M$ and the collapsing and evaporating semiclassical geometry. We will fix a given geometry which includes the effects of the slow evaporation and we consider quantum fluctuations around this state that do not change the geometry in any important fashion. In particular we could assume that all quantum fluctuations around this state do not change the energy of the state.

The net result is that the initial matter state $|\psi\rangle_M$ evolves into a state in $H_M \otimes H_{in} \otimes H_{out}$ given roughly by $|\psi\rangle_M |U\rangle_{in \times out}$. One can view Hawking radiation as the pair creation of a positive energy particle that goes to infinity and a negative energy particle that falls into the black hole. The pair is created in a particular entangled state. So the Unruh state can be viewed as an entangled thermal state. For our purposes it is more convenient to view it in the microcannonical ensemble considering all states in $H_{out}$ that have fixed energy equal to the black hole mass. So we write

$$|U\rangle = \sum_i \frac{1}{\sqrt{N}} |i\rangle_{in} |i\rangle_{out} \qquad (2.1)$$

where $N$ is the dimension of these Hilbert spaces, i.e., $N = e^S$, with $S$ the entropy of the black hole. We are describing the black hole interior by the tensor product of two Hilbert spaces $H_M \otimes H_{in}$. The main distinction between them is the following. The black hole symmetry $\partial/\partial t$ is spacelike inside the horizon so states can have both positive and negative eigenvalues. States with positive eigenvalue describe the collapsing matter and belong to $H_M$, while states with negative eigenvalue belong to the Hilbert space of the infalling Hawking partners $H_{in}$. Strictly speaking this separation is only approximate since particles interact and also the evaporating black hole geometry does not have an exact Killing field $\partial/\partial t$. If we trace over the unobserved interior, we get a density matrix in $H_{out}$.

## 2.2. Review of quantum mechanics with final state boundary conditions

In quantum mechanics, where the probabilities are computed as the square of amplitudes, one cannot assign a probability to every history. The usual probability sum rules would not be satisfied due to interference. The traditional response is to say that one can only assign probabilities to histories that are measured. The measurement process is supposed to "collapse the wave function" and remove the interference. This is not applicable to a complete description where any measuring device must be part of the quantum state. Moreover, it would not be applicable near singularities where there are no observers around to do any measurements. A more general statement about when probabilities can be assigned is in terms of decoherence.

In a path integral description, the (fine grained) histories are the individual paths. Coarse grained histories are collections of paths satisfying specified conditions. It will be more convenient for us to adopt a different description in terms of projection operators. Alternatives at a moment of time $t_k$ can be represented by a set of orthogonal projections $\{P_{\alpha_k}(t_k)\}$ with $\sum_{\alpha_k} P_{\alpha_k}(t_k) = 1$. A set of histories is defined by giving a series of such alternatives at a sequence of times $t_1, \cdots, t_n$. An individual history is a sequence of alternatives $(\alpha_1, \cdots, \alpha_n) = \alpha$, and is denoted by the corresponding chain of projections

$$C_\alpha \equiv P_{\alpha_n}(t_n) \cdots P_{\alpha_1}(t_1) \tag{2.2}$$

The decoherence functional is

$$D(\alpha, \alpha') = Tr[C_\alpha \rho_i C_{\alpha'}^\dagger] \tag{2.3}$$

where $\rho_i$ denotes the initial state, which in general could be a density matrix. When the off diagonal elements of the decoherence functional vanish (or are negligibly small) the set of histories is said to decohere and are sometimes called "consistent histories". In this case, there is no interference between pairs of distinct histories and the probability of an individual history is just the diagonal element $p(\alpha) = D(\alpha, \alpha)$. These probabilities sum to one since $\sum_{\alpha,\alpha'} D(\alpha, \alpha') = 1$ and the off diagonal elements do not contribute.

This is just a reformulation of ordinary quantum mechanics. It is manifestly time asymmetric, since there is an initial state $\rho_i$ specified, but no final state. This is the usual situation in quantum mechanics. When we do computations we sometimes specify an initial and a final state to compute an amplitude, or a matrix element of an $S$ matrix. But when we use these answers we usually sum over all possible final states. The idea of having a more time "symmetric" quantum mechanics description was explored in [6,7,8,9]. In this modification, one simply adds a final density matrix into the decoherence functional

$$D(\alpha, \alpha') = \mathcal{N} Tr[\rho_f C_\alpha \rho_i C_{\alpha'}^\dagger] \tag{2.4}$$

where

$$\mathcal{N}^{-1} = Tr[\rho_f \rho_i] \tag{2.5}$$

The condition for when probabilities can be defined is again the vanishing of the off diagonal elements of $D$.

## 3. A proposal for the black hole singularity

### 3.1. Proposal for a final state

If quantum gravity is an ordinary quantum theory for observers outside the event horizon, it should lead to unitary evolution. In particular, we expect that black hole evaporation as seen by an asymptotic observer is a unitary process. In other words, in these quantum gravity theories one has a unitary $S$ matrix, $S : H_M \to H_{out}$, that determines what comes out from a black hole. In AdS/CFT one would compute this $S$ matrix by doing computations in the unitary CFT. In order to agree with Hawking's answer we want to say that a suitable average of this $S$ matrix gives rise to an approximately thermal state. In explicit computations one finds that this is the case [14].

Still one would like to understand how to modify Hawking's original computation so that it is consistent with unitarity. Information loss comes from the fact that particles fall into the singularity and we need to sum over all the possible final quantum states in the interior. In other words, in the standard treatment the singularity looks like another unobserved "asymptotic" region. Here we propose that there is a final state boundary condition at the singularity.

The first part of Hawking's computation is the same as before. We start with a state $|\psi\rangle_M \in H_M$ describing the pure state that will form the black hole. In a Heisenberg representation (implicit in the time dependent projection operators used in the previous section) the initial vacuum state of the fluctuations can be reexpressed as the Unruh state (2.1), so we have

$$|\psi\rangle_M |U\rangle_{in \times out} = |\psi\rangle_M \times \sum_i \frac{1}{\sqrt{N}} |i\rangle_{in} |i\rangle_{out} \equiv |\Psi\rangle \tag{3.1}$$

Now we propose that at the singularity we have a final state boundary condition, given by a state called $\langle BH|$, which lives in the Hilbert space describing the inside of the black hole, $H_M \otimes H_{in}$,

$$\langle BH| = \frac{1}{N} \sum_{m,i} S_{mi} \langle m|_M \langle i|_{in} \qquad (3.2)$$

We will further demand that the matrix $S$ is unitary. In the natural Fock space basis associated with Hawking particles and infalling matter, this matrix $S$ will be very complicated, essentially a completely random matrix. In particular, this matrix would *not* result from a local boundary condition on the fields at the singularity, since this would give us a relatively simple matrix in the usual Fock basis. We also expect that this matrix $S$ is non-local in the coordinates along the singularity. A complete theory of quantum gravity would let us compute it.[2]

In terms of the decoherence functional (2.4), the initial density matrix is $\rho_i = |\Psi\rangle\langle\Psi|$ and the final density matrix is

$$\rho_f = |BH\rangle\langle BH| \times \frac{1}{N} I_{out}. \qquad (3.3)$$

Now consider only projection operators $P^{out}$ associated with an outside observer, and let $C_\alpha^{out}$ be a chain of such projections. Then the decoherence functional (2.4) reduces to

$$D(\alpha, \alpha') = Tr[C_\alpha^{out} \tilde{\rho} C_{\alpha'}^{\dagger out}] \qquad (3.4)$$

where $\tilde{\rho} = |\tilde\Psi\rangle\langle\tilde\Psi|$ with

$$|\tilde\Psi\rangle = N^{3/2} \langle BH|\psi\rangle_M |U\rangle_{in \times out} = S_{mi} \psi_M^m |i\rangle_{out} \qquad (3.5)$$

and $\psi_M^m = {}_M\langle m|\psi\rangle_M$. This is equivalent to unitary evolution from $|\Psi\rangle$ to $|\tilde\Psi\rangle$. Furthermore, since the matrix $S$ is a random matrix, the final state looks thermal if we average over many matrix elements, i.e. sufficiently coarse grained observables are thermal. More explicitly, suppose we write the out Hilbert space as $H_{out} = H_1 \otimes H_2$ and the index $i = (i_1, i_2)$. If we have an observable that only acts on $H_1$ but does not act on $H_2$ then we can define a density matrix by summing over all states in $H_2$. This leads to

$$\tilde{\rho}_{i_1 i_1'} = \sum_{i_2} \psi_M^{*m'} S^*_{m' i_1 i_2} S_{m i_1' i_2} \psi_M^m \sim \delta_{i_1 i_1'} \qquad (3.6)$$

where we have used that $S$ is a random matrix. So we see that this is the expected density matrix for a thermal state in the microcannonical ensemble.

---

[2] The fact that the classical evolution approaching a singularity is chaotic [15] suggests that the quantum evolution operator will effectively be a random matrix. It is tempting to speculate that there may be a connection between this random dynamics and the random state at the singularity.

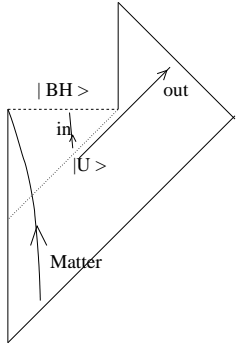

**Fig. 1:** Penrose diagram for a black hole that forms and evaporates. We see the matter falling in. We also see the pairs that are created at the horizon. One element of the pair goes to infinity the other falls into the black hole. The black hole final boundary condition ensures that everything that is falling into the black hole annihilates completely leaving nothing behind.

Of course the state $|BH\rangle$ carries zero charge for all gauge charges. These charges are conserved by $|BH\rangle$. It is believed that these are the only conserved charges in a theory of quantum gravity.

Some people have speculated that there might be another semi-classical region of spacetime after the black hole singularity (see, e.g., [16]). In their view, the singularity is a mere artifact of the semi-classical approximation. Our proposal is incompatible with the idea that there is a second asymptotic region beyond the singularity. More precisely, our ideas are incompatible with any information transfer between the black hole and a second hypothetical asymptotic region beyond the singularity.

Our proposal of a final state is very closely connected to the proposal in [17] where the interior of the black holes are treated as wormholes [18]. Then if we consider the theory for fixed $\alpha$ parameters, the wavefunction for the interior of black holes is fixed.

Finally we should point out that the assumption of unitarity for the matrix $S$ is a very strong assumption. In particular, in the presence of interactions that mix $H_M$ and $H_{in}$ one would need to adjust $S$ so that in the end we effectively have a unitary matrix[3].

*3.2. Arrow of time*

When one has a final boundary condition one might think that the arrow of time would be reversed. Notice however that a local observer falling into a black hole has access to a very small subset of the Hilbert space $H_M \otimes H_{in}$. Namely, any experiment he does has a finite number of possible practically distinguishable outcomes, since he has finite energy and finite time to do the experiment. This finite number of outcomes is much smaller than the black hole entropy, which is the dimension of the Hilbert spaces we are talking about. This implies that in any measurement the infalling observer would need to average over unobserved quantities. In particular, let us assume that he does essentially no observation involving the $H_{in}$ degrees of freedom. Then using (3.3), the above formalism would involve an effective density matrix

$$\hat{\rho}_f = Tr_{H_{in}, H_{out}} \rho_f = \frac{1}{N} \sum_m |m\rangle\langle m| \qquad (3.7)$$

---

[3] This point has recently been stressed by D. Gottesman and J. Preskill [19]. We thank them for correspondence on this issue.

So we end up with usual quantum mechanics where the final density matrix is complete ignorance, for observables involving only $H_M$. Of course more realistic observables will involve some degrees of freedom from both $H_M$ and $H_{in}$, but we will still need to sum over a large fraction of the states in $H_{in}$ which will produce a density matrix in the subspace of accessible final states for the experiment which will be close to the identity for his possible accessible final states.

Note that when we solve the classical equations for collapse to a black hole we are neglecting any information about $H_{in}$, namely the precise quantum state of the Hawking quanta. So the above argument implies that it is reasonable to treat the singularity as capable of "absorbing" any classical solution, which is the standard assumption. It is crucial that the final state boundary condition is a condition on the quantum state which contains a superposition of many macroscopically distinct states[4].

*3.3. Complementarity*

Another solution that has been proposed for the black hole information problem is that observables in the interior and the exterior of the black hole do not commute [21]. This is known as black hole complementarity. In this approach, the horizon is treated as a special surface. In the description we propose here the unexpected physics is happening at the singularity. We have seen that probabilities can only be assigned to sets of histories that decohere. There is a sense in which decoherence implies complementarity. In order to talk about observations both outside and inside the black hole we need to construct an appropriate set of histories that decohere. A local measurement in the interior will not generically decohere with a measurement in the exterior due to the final state boundary condition.

---

[4] The idea of a final boundary condition in a more classical sense was proposed in [20], and it lead to a reversal of the arrow of time in the interior.

Let us examine this in more detail. We will be interested in doing measurements in Hilbert spaces $H_M$, $H_{in}$ and $H_{out}$ via projection operators $P^M$, $P^{in}$ and $P^{out}$ which act on the respective Hilbert spaces. Let us introduce a matrix $U_{ij}$ characterizing Unruh's state, $|U\rangle = U_{ij}|i\rangle_{in}|j\rangle_{out}$ (In the microcannonical ensemble $U_{ij} \sim \delta_{ij}$). On the state $|\psi\rangle_M|U\rangle$ these projection operators give the state

$$|m\rangle_M|i\rangle_{in}|j\rangle_{out} P^M_{mm'}\psi_{M\ m'} P^{in}_{ii'} P^{out}_{jj'} U_{i'j'} \tag{3.8}$$

where $P_{ii'} = \langle i|P|i'\rangle$, etc. After imposing the final state boundary condition (3.2) we basically "sew" together $H_{in}$ and $H_M$ so we are left with a state in the out Hilbert space proportional to

$$|j\rangle_{out} S_{mi} P^M_{mm'}\psi_{M\ m'} P^{in}_{ii'} P^{out}_{jj'} U_{i'j'} \tag{3.9}$$

In terms of the matrices $S$ and $U$ we can write this state as

$$P^{out} U^t (P^{in})^t S^t P^M |\psi_M\rangle \tag{3.10}$$

Note that in this expression we have the operators in the Matter, In and Out Hilbert space in a form which suggests that we should think of a time ordering sequence where the matter particles "reflect" from the singularity, they travel backwards in time as particles in $H_{in}$ and then out to $H_{out}$ thanks to the entanglement in Unruh's state. This reflects the basic flow of information we would have in the system.

Note also that natural observables given by $P^{out}$ would not decohere with natural local observables given by $P^M$ due to the fact that $S$ is a random matrix that will make the corresponding operators not commute in (3.10). Of course we can choose projection operators in $H_M$ and $H_{out}$ that will effectively commute in (3.10), but those operators will not be natural local operators in the respective regions.

## 4. Generalizations

We have confined the above discussion to an evaporating black hole with a spacelike curvature singularity such as a Schwarzschild black hole. The main point of the paper is to point out that in order for unitary to be preserved, we should prevent the transfer of the information carried by the infalling matter to a second asymptotic region inside the black hole. (The "first" asymptotic region is the one outside the black hole). In the standard Hawking computation the interior is effectively treated as a second unobserved asymptotic region. In this section we discuss generalizations of this idea to other black holes and cosmological singularities.

*4.1. Other black holes*

In charged or rotating black holes the singularity is timelike, and there is a second asymptotically flat region that can be accessed by going behind the horizon. In fact, the Reissner-Nordstrom metric is timelike geodesically complete. In a theory with particles satisfying $|q| > m$, the black hole preferentially radiates away its charge and evaporates completely. Similarly, a rotating black hole will radiate away its angular momentum [22]. In order for the evaporation to be unitary, no information should pass through the black hole. Thus, we expect that in these cases, the full semiclassical evolution will result in a curvature singularity cutting off spacetime inside the horizon. We would then impose final boundary conditions at this singularity similar to the Schwarzschild case. This singularity could be produced by instabilities of the inner Cauchy horizon, and might be null in certain regions.

The evaporation of a charged black hole in a theory with all particles satisfying $|q| \leq m$ is expected to result in an extremal black hole. In this case, since the black hole does not disappear completely, the final Hilbert space is a product of the usual out states $H_{out}$ and the states of the extremal black hole $H_{ext}$. The extreme Reissner-Nordstrom metric also contains Cauchy horizons. Unitarity suggests that in the process of evaporation down to extremality, the Cauchy horizon will again become singular and final state boundary conditions should be imposed.

Large black holes in an asymptotically AdS spacetime do not evaporate. They exist in a type of equilibrium where the Hawking radiation is reabsorbed at the same rate that it is emitted. We believe that there will also be a final state boundary condition at the black hole singularity. One difference with the evaporating case is that the space near the singularity now has infinite volume. It is not effectively compact. Nevertheless we expect that the final boundary state still has some "entropy" of order the black hole entropy. This deserves further study.

The three dimensional BTZ black hole [23] is also eternal. It has positive specific heat and will come to equilibrium with its Hawking radiation. Classically, the geometry is locally $AdS_3$ except in some spacelike region where it has a compactified Milne type singularity (in the non-rotating case). In the rotating case the spacetime is locally $AdS_3$ everywhere but there are closed null (and timelike) curves. As above, we suspect that in a full semi-classical calculation, there will be a curvature singularity and final boundary conditions should be imposed.

*4.2. Cosmological singularities*

Some recent proposals for dealing with the big bang singularity, such as the Ekpyrotic or cyclic models [24,25] involve singularities which are similar to black hole singularities. Namely, near the bounce region of these models the dynamics is five dimensional. The extra dimension together with time describe a compactified Milne region near the singularity [26]. Precisely the same type of singularity occurs for the (non-rotating) three dimensional BTZ black hole. So our previous discussion suggests that no information about the state of the system, including information about quantum fluctuations, can be transferred from the region before the bounce to the region that comes after the bounce.

It is natural to extrapolate from this that all cosmological singularities will have the property that no information can pass through[5]. In other words, bouncing universes such as the those in the pre-big bang scenario [27], probably do not exist. Instead we have initial and final state boundary conditions. The most well known example of an initial state boundary condition is the Hartle-Hawking no-boundary proposal [11].

*4.3. AdS/CFT and the wavefunction of the universe*

The euclidean version of the AdS/CFT duality provides support for the idea that there is a unique wavefunction for compact universes and it also provides an exact way to compute it in a particular limit. In this section we explain this statement in more detail[6].

A naive canonical quantization of gravity suggest the existence of a wavefunction of the universe which obeys the Wheeler de Witt equation. We can think of this wavefunction as a function of the spatial geometry. We are interested in compact spatial geometries, such as a sphere. It is convenient to write the metric of the $d$-dimensional spatial sections as $ds^2 = a^2 d\hat{s}^2$, where $d\hat{s}^2$ is normalized to have unit volume and $a$ is the overall scale factor. Then the wavefunction becomes a function of $a$ and the rescaled spatial geometry $\hat{g}$. When we solve the Wheeler de Witt equation we might find that the wavefunction $\Psi(a, \hat{g})$ is either an oscillatory function of $a$ or a real exponentially increasing or decreasing function of $a$. The oscillatory pieces of the wavefunction originate in Lorentzian geometries, while the exponentially increasing or decreasing pieces are associated to Euclidean geometries. So the same gravitational Lagrangian can lead to both exponential or oscillatory behavior, depending on the values of $a$ and the geometry of the spatial section, the values of scalar fields fields, etc .

---

[5] Our arguments are clearer for compact universes. In the noncompact case, new subtleties might arise.

[6] Lorentzian AdS/CFT, like strings in asymptotically flat spacetime, allow arbitrary initial states.

In principle this equation can have many solutions. In [11] it was proposed that one should pick the wavefunction by doing a functional integral over all $d+1$ dimensional geometries with no other boundary. A problem with this proposal is that if one weights each geometry by the standard Einstein action, it is not well defined even perturbatively, due to nonrenormalizability. Furthermore, we do not know if the concept of a wavefunction of the universe makes sense non-perturbatively in quantum gravity.

Now, consider a $d+1$ dimensional gravity theory with a negative cosmological constant. In this case, for large $a$, one can find Euclidean geometries that locally behave as

$$ds^2 = a^2(\rho) d\hat{s}_d^2 + d\rho^2 , \qquad a \sim e^\rho \tag{4.1}$$

Such geometries lead to WKB wavefunctions $\psi \sim e^{-S_E}$ and the Euclidean action goes as

$$S_E = -\int_{\rho < \rho_{max}} \sqrt{G}(R - 2\Lambda) - 2\oint K \propto -a_0^d \tag{4.2}$$

with $a(\rho_{max}) = a_0$. This leads to an approximate wavefunction of the form

$$\psi = e^{-S_{loc}(a,\hat{g})} \mathcal{A}(a, \hat{g}) \tag{4.3}$$

where $S_{loc}$ a piece which is local in the boundary metric and diverges when we move into the asymptotic $AdS$ region. It has the form $S_{loc} \sim c_0 a_0^d \int \sqrt{\hat{g}} + c_1 a_0^{d-2} \int \sqrt{\hat{g}} \hat{R} + \cdots$, with $c_i$ some constants [28]. The expansion terminates in the $[d/2]$th term. After extracting this local divergent piece we find that $\mathcal{A}$ has a finite limit at $a \to \infty$. This limiting value is conjectured to be equal to the partition function of a conformal field theory on the geometry given by $\hat{g}$ [3,4]

$$\lim_{a \to \infty} \mathcal{A}(a, \hat{g}) = \mathcal{Z}(\hat{g}) \tag{4.4}$$

This relationship is conjectured to be valid non-perturbatively on the two sides. So the AdS/CFT conjecture gives a way to compute the wavefunction of the universe in a non-perturbative way in the limit that $a$ is very large. In the leading semiclassical approximation it agrees with the Hartle-Hawking no boundary proposal. This AdS/CFT prescription actually includes some geometries which are not smooth. Geometries which are singular in a supergravity approximation are allowed as long as they are nonsingular in the full string theory.

Note that the precise Lagrangian of the CFT is conjectured to depend only on the boundary conditions for the metric and all other fields at infinity. In other words, the data that is necessary for determining the CFT is precisely the same data that the wavefunction of the universe is supposed to depend on in this limit.

This is evidence that the wavefunction of the universe is unique. If we had different choices for the wavefunction of the universe then we would have different possible values for the partition function of the same CFT. This would imply that the field theory is not well defined. Note that there could be many geometries that contribute to a given wavefunction [4]. This unique wavefunction contains a huge number of large semiclassical geometries.[7]

Of course, what we have said so far applies to the part of the wavefunction that correspond to Euclidean geometries. But it is reasonable to think that the uniqueness of the wavefunction in these exponentially varying regions also leads to a unique solution in the Lorentzian part, as originally argued in [11].

This uniqueness of the wavefunction for compact Lorentzian geometries is intimately connected with our proposal for the black hole final state. In fact the spatial sections for the interior of an evaporating black hole are effectively compact, since the Schwarzschild "time" direction is finite if the black hole evaporates in finite time.

The implications of AdS/CFT to the unitarity problem were also explored in [29].

---

[7] In fact, it seems likely to imagine that all possible boundary conditions correspond to all possible field theories we can define. Most of them will not correspond to geometries that are large in Planck or string units.

## 5. Conclusion

We have suggested a possible resolution of the apparent contradiction between string theory and semiclassical arguments over whether black hole evaporation is unitary. By imposing a final state boundary condition at the black hole singularity, one circumvents the usual causality problem and obtains unitary evolution even in a semiclassical treatment.

The ideas in this paper are completely consistent with the ideas of Hartle and Hawking about a unique wavefunction for the universe [11]. It is precisely this uniqueness that saves the day in black hole evaporation. Notice that a unique initial quantum state does not imply a unique macroscopic state, we could have a superposition of different macrocopically distinct universes.

Our proposal is clearly very speculative, and we have not given any constructive method for computing the black hole final state. In principle, this can be deduced from a precise calculation of the evolution $|\psi\rangle_M \to |\psi\rangle_{out}$. Hopefully, we will soon discover some techniques to compute it!

## Acknowledgments

It is a pleasure to thank J. Hartle and J. Polchinski for discussions. This work was supported in part by NSF grants PHY-0244764 and DOE grant DE-FG02-90ER40542.